\documentclass{JHEP3}
\usepackage{epsfig}
\usepackage{graphicx}
\usepackage{amssymb}
\usepackage{latexsym}

\title{Dynamical evolution of phantom scalar perturbation
in the Schwarzschild black string spacetime}
\author{Songbai Chen \\
Institute of Physics and Department of Physics,
Hunan Normal University,  Changsha, Hunan 410081, P. R. China \\
Key Laboratory of Low Dimensional Quantum Structures and Quantum
Control (Hunan Normal University), Ministry of Education, P. R.
China. E-mail: \email{csb3752@163.com }}
\author{Jiliang Jing  \\
Institute of Physics and Department of Physics,
Hunan Normal University,  Changsha, Hunan 410081, P. R. China \\
Key Laboratory of Low Dimensional Quantum Structures and Quantum
Control (Hunan Normal University), Ministry of Education, P. R.
China. E-mail: \email{jljing@hunnu.edu.cn}}

\abstract{
Using Leaver's continue fraction and time domain method,
we study the wave dynamics of phantom scalar perturbation in a
Schwarzschild black string spacetime.  We find that the quasinormal
modes contain the imprint from the wavenumber $k$ of the fifth
dimension. The late-time behaviors are dominated by the difference
between the wavenumber $k$ and the mass $\mu$ of the phantom scalar
perturbation. For $k<\mu$, the phantom scalar perturbation in the
late-time evolution grows with an exponential rate as in the
four-dimensional Schwarzschild black hole spacetime. While, for
$k=\mu$, the late-time behavior has the same form as that of the
massless scalar field perturbation in the background of a black
hole. Furthermore, for $k>\mu$, the late-time evolution of phantom
scalar perturbation is dominated by a decaying tail with an
oscillation which is consistent with that of the usual massive
scalar field. Thus, the Schwarzschild black string is unstable only
against the phantom scalar perturbations which satisfy the
wavelength $\lambda>2\pi/\mu$. These information can help us know
more about the wave dynamics of phantom scalar perturbation and the
properties of black string.
}

 \preprint{2008/12/14}

\begin{document}
\section{Introduction}

The current observations confirm that the expansion of the present
Universe is speeding up rather than slowing down. This may indicate
that our Universe contains dark energy which is an exotic energy
component with negative pressure and  constitutes about $72\%$ of
present total cosmic energy. The leading interpretation of such a
dark energy is a cosmological constant with equation of state
$\omega_x=-1$ \cite{1a}. Although the cosmological constant model is
consistent with observational data, at the fundamental level it
fails to be convincing. The vacuum energy density falls far below
the value predicted by any sensible quantum field theory, and it
unavoidably yields the coincidence problem, namely, ``why are the
vacuum and matter energy densities of precisely the same order
today?". Therefore the dynamical scalar fields, such as quintessence
\cite{2a}, k-essence \cite{3a} and phantom field \cite{4a}, have
been put forth as an alternative of dark energy.

Comparing with other dynamical scalar fields, the phantom field
model is more interesting  because that it has a negative kinetic
energy which can be appeared in the quantum particle creation
processes in curved backgrounds \cite{41a} or can be motivated from
S-brane constructs in string theory \cite{42a}. The presence of such
a negative kinetic energy leads to that the null energy condition is
violated and the equation of state $\omega_x$ of the phantom energy
is less than $-1$. In the Einstein cosmology, the dynamical
evolutions of phantom field \cite{6a1,6a11,6a2,6a3,6a4} tell us that
its energy density increases with the time and approaches to
infinity in a finite time. In other words, the Universe dominated by
phantom energy will blow up incessantly and arrive at a big rip
finally, which is a future singularity with a strong exclusive force
so that anything in the Universe including the large galaxies will
be torn up. The thermodynamical properties of a phantom Universe are
strange and such the universe has a negative entropy diverging near
the big rip. Although the phantom energy possesses such exotic
properties, it is favored by recent precise observational data
involving CMB, Hubble Space Telescope and type Ia Supernova
\cite{5a}.

In the background of black hole spactimes, phantom field also
exhibits some peculiar properties. E. Babichev and his co-workers
\cite{Eph} found that all black holes in the phantom universe lose
their masses to vanish exactly in the big rip. Therefore, it is
possible that as a charged black hole absorbs the phantom energy the
charge of a black hole will be larger than its mass. This is a
serious threat to the cosmic censorship conjecture \cite{Eph1,Eph2}
. In our previous paper \cite{sb1}, we studied that the wave
dynamics of the phantom scalar perturbation in the Schwarzschild
black hole spacetime. Our result shows that in the late-time
evolution the phantom scalar perturbation grows with an exponential
rate rather than decays as the usual scalar perturbations
\cite{qn1,qn2,qn3,ml1,ml2}. Moreover we also find that the phantom
scalar emission will enhance the Hawking radiation of a black hole
\cite{sb2}. These results could help us to get a deeper understand
about dark energy and black hole physics.

The discussions above tell us that comparing with other fields the
phantom field plays the entirely different roles both in the
cosmology and black hole physics. The recent investigations else
show that the peculiar properties of phantom energy make it possible
to be a candidate for exotic matter to construct wormholes
\cite{wmh}. However, all of the above investigations of phantom
fields do not consider the case of black string. Therefore, what
effects of the phantom field on black strings is still open. It is
well known that black string is an important kind of black objects
in the high dimensional gravity theories. In general, such a
Schwarzschild black string suffers from the so-called
Gregory-Laflamme instability, namely, long wavelength gravitational
instability of the scalar type of the metric perturbations
\cite{GF1,GF2}. The Gregory-Laflamme instability has been
extensively studied in the last decade \cite{GF3,GF4} and the
threshold values of the wavenumber $k$ at which the instability
appears are obtained \cite{GF5}. For the usual scalar,
electromagnetic and Dirac field perturbations, the Schwarzschild
black string does not meet such a Gregory-Laflamme instability
problem. Since the phantom field presents many peculiar behaviors in
many fields, it is of interesting to study its dynamical evolution
in the black string spacetime and to probe whether it presents some
new behaviors. The main purpose of this paper is to investigate the
quasinormal modes and late-time behavior of the phantom scalar
perturbation in the Schwarzschild black string background through
the well-known continue fraction \cite{Leaver} and time domain
methods \cite{T1}, and to discuss further the problems of
instability of black string and a phantom scalar field.

The paper is organized as follows: in the following section we give
the wave equation of phantom scalar field in the Schwarzschild black
string spacetime. In Sec.III, we calculate the fundamental
quasinormal modes of the phantom scalar perturbation by the continue
fraction technique and plot the late-time evolution by time domain
method. Finally in the last section we include our conclusions.

\section{The wave equation of phantom scalar field in the Schwarzschild black string spacetime}

Let us now to consider a Schwarzschild black string spacetime with
the compact fifth $z$-coordinate, whose metric in the standard
coordinate can be described by
\begin{eqnarray}
ds^2=(1-\frac{2M}{r})dt^2-(1-\frac{2M}{r})^{-1}dr^2-r^2(d\theta^2+\sin^2\theta
d\phi^2)-dz^2.\label{bstr1}
\end{eqnarray}
The horizon event horizon is situated at $r=2M$ and we assume that
the $z-$direction is periodically identified by the relation
$z=z+2\pi R$.

In the background spacetime (\ref{bstr1}), the action of the phantom
scalar perturbation with the negative kinetic energy term can be
expressed as \cite{4a, phso1}
\begin{eqnarray}
S=\int d^4xdz \sqrt{g}\bigg[-\frac{R}{16\pi
G}-\frac{1}{2}\partial_{\mu}\psi\partial^{\mu}\psi+V(\psi)\bigg].
\end{eqnarray}
Here we take metric signature $(+----)$. The usual ``Mexican hat"
symmetry breaking potential has the form
\begin{eqnarray}
V(\psi)=-\frac{1}{2}\mu^2\psi^2+\frac{\kappa}{4}\psi^4,
\end{eqnarray}
where $\mu$ is the mass of the scalar field and $\kappa$ is the
coupling constant. Here we treat the phantom scalar field as an
external perturbation and suppose it does not change the metric of
the background. As in Refs. \cite{sb1,sb2}, we only consider the
case $\kappa=0$ for conveniently. Thus, varying the action with
respect to $\psi$, we obtain the wave equation for phantom scalar
field in the curve spacetime
\begin{eqnarray}
\frac{1}{\sqrt{g}}\partial_{\mu}(\sqrt{g}g^{\mu\nu}\partial_{\nu})
\psi-\mu^2\psi=0.\label{WE}
\end{eqnarray}
Substituting Eq. (\ref{bstr1}) into Eq. (\ref{WE}), and separating
variable
\begin{eqnarray}
\psi=e^{-i\omega
t}e^{ikz}\frac{R(r)Y_{lm}(\theta,\phi)}{r},~~~~~k=\frac{N}{R},
~~~~~N\in Z, \label{NRR}
\end{eqnarray}
we can obtain the radial equation for the scalar perturbation in the
Schwarzschild black string spacetime
\begin{eqnarray}
\frac{d^2 P(r)}{dr^2_*}+[\omega^2-V(r)]R(r)=0,\label{jw}
\end{eqnarray}
where $r_*$ is the tortoise coordinate (which is defined by
$dr_*=\frac{r}{r-2M}dr$) and the effective potential $V(r)$ reads
\begin{eqnarray}
V(r)=\bigg(1-\frac{2M}{r}\bigg)\bigg(\frac{l(l+1)}{r^2}+\frac{2M}{r^3}+k^2-\mu^2\bigg).\label{efp}
\end{eqnarray}
The form of the effective potential $V(r)$ (\ref{efp}) for the
phantom scalar perturbation is similar to that in the Schwarzschild
black hole spacetime \cite{sb1}, but there is an additional "mass"
term $k^2$ in $V(r)$, which originates from the compact fifth
dimension of a Schwarzschild black string. The presence of the term
$k^2$ changes the behavior of the effective potential $V(r)$ of the
phantom scalar field. With the increase of $k$, the peak height of
the potential barrier increases. Moreover, at the spatial infinity
the effective potential $V(r)$ approaches to a constant $k^2-\mu^2$
rather than $-\mu^2$ in Schwarzschild black string spacetime. For
the case $k^2$ is smaller than $\mu^2$, we find that $V(r)$ is
negative as $r$ tends to infinity. It is similar to that in
Schwarzschild black hole spacetime in which the negative limit
$V(r)|_{r\rightarrow \infty}$ yields that the phantom scalar
perturbation in the late-time evolution grows with an exponential
rate. However, for $k^2$ is larger than or equal to $\mu^2$, the
effective potential $V(r)$ (\ref{efp}) is no more negative again in
the physical regime. In general, the positive potential means that
the background spacetime is stable against an external perturbation.
This implies also that the evolution of phantom scalar perturbation
possesses some new properties in the Schwarzschild black string
spacetime.

\FIGURE[!t]{ \centerline{
\includegraphics[width=0.4\textwidth,angle=0]{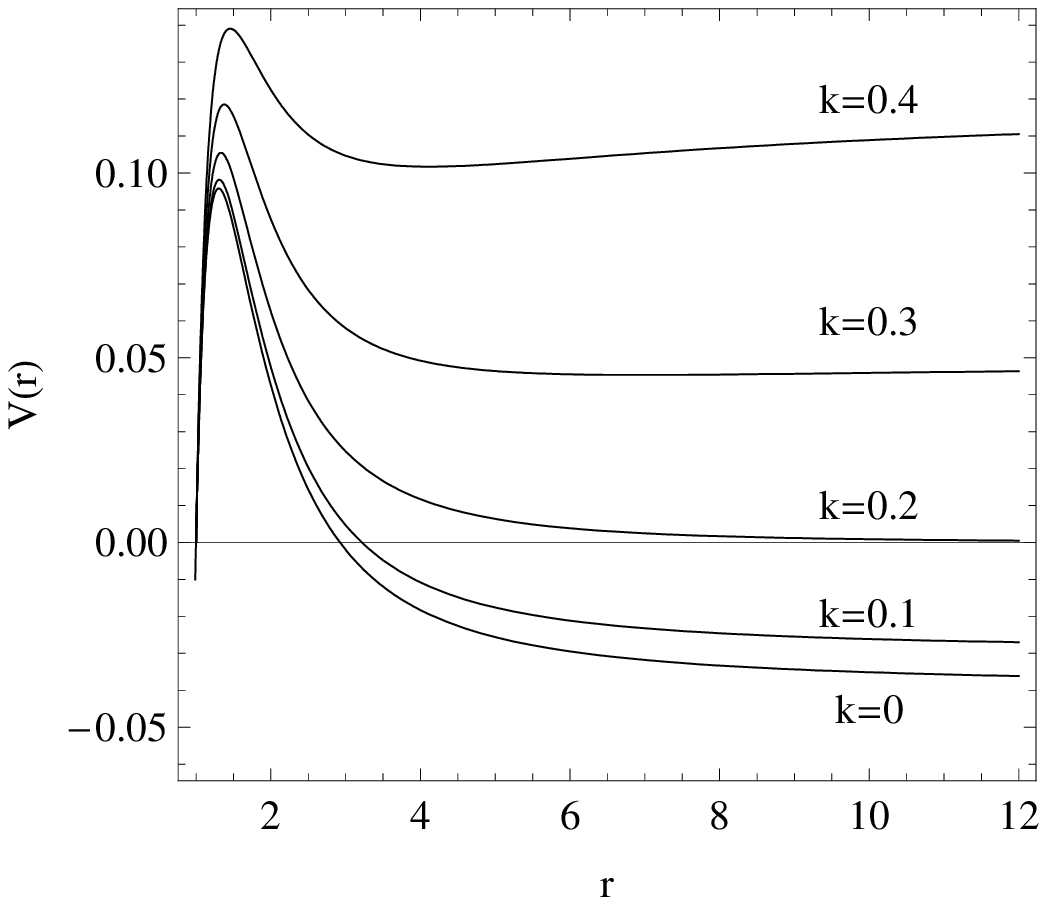}~~\includegraphics[width=0.4\textwidth,angle=0]{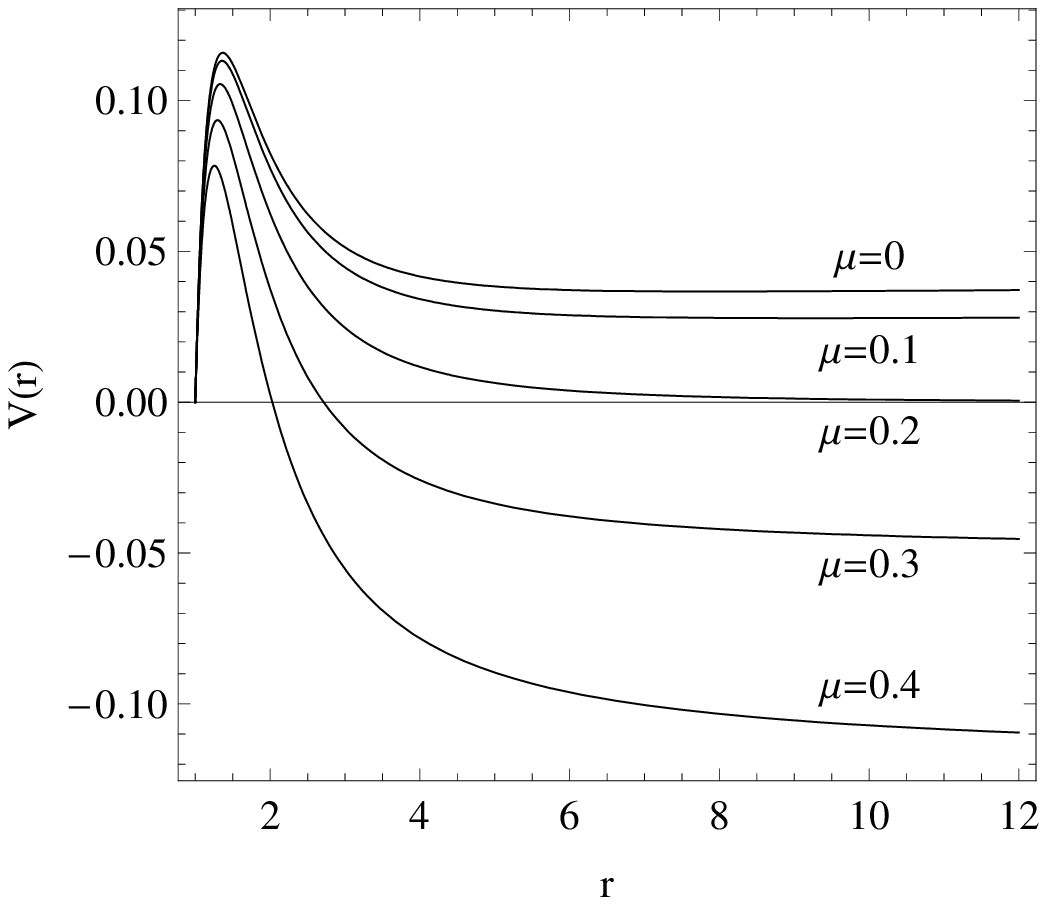}
\caption{Variety of the effective potential $V(r)$ of phantom scalar
perturbation with $r$. The left for fixed $\mu=0.2$ and the right
for fixed $k=0.2$ in the Schwarzschild black string spacetime. Here
$l=0$ and $M=0.5$. \label{fig1}} }}

\section{Evolution of phantom scalar perturbation in the Schwarzschild string spacetime}

In this section we will study the wave dynamics of phantom scalar
perturbation in the Schwarzschild string spacetime by using Leaver's
continue fraction \cite{Leaver} and time domain method \cite{T1}.

Let us now to calculate the fundamental quasinormal modes which are
dominated in the late time oscillations. The boundary conditions on
the wave function $R(r)$ of phantom scalar perturbation in the
Schwarzschild black string spacetime can be expressed as
\begin{eqnarray}
R(r)=\bigg\{\begin{array}{lll}
(r-1)^{-i\omega},~~~~~~~~~~~~r\rightarrow 1,\\
\\
r^{\frac{i(\chi^2+\omega^2)}{2\chi}}e^{i\chi
r},~~~~~~~~~r\rightarrow \infty,
\end{array}\label{bd}
\end{eqnarray}
where we set $2M=1$ and $\chi=\sqrt{\omega^2-k^2+\mu^2}$. A solution
to Eq.(\ref{jw}) that has the desired behavior at the boundary can
be written as
\begin{eqnarray}
R(r)=r^{i(\chi+\omega)-\frac{i\mu^2}{2\chi}}(r-1)^{-i\omega}e^{i\chi
r}\sum_{n=0}^{\infty}a_n\bigg(\frac{r-1}{r}\bigg)^n.\label{wbes}
\end{eqnarray}
Substituting (\ref{wbes}) into (\ref{jw}), we find that the
coefficients $a_n$ of the expansion satisfy a three-term recurrence
relation staring from $a_0=1$
\begin{eqnarray}
&& \alpha_0 a_1+\beta_0a_0=0,\nonumber\\
&& \alpha_na_{n+1}+\beta_na_n+\gamma_na_{n-1}=0,~~~~~~n=1,~2,~
\ldots. \label{wbess}
\end{eqnarray}
In terms of $n$ and the black hole parameters, one find that the
recurrence coefficients $\alpha_n$, $\beta_n$ and $\gamma_n$ have
the forms
\begin{eqnarray}
&& \alpha_n=n^2+(C_0+1)n+C_0,\nonumber\\
&& \beta_n=-2n^2+(C_1+2)n+C_3,\nonumber\\
&& \gamma_n=n^2+(C_2-3)n+C_4-C_2+2,
\end{eqnarray}
where the intermediate constants $C_n$ are the functions of the
variables $\omega$ and $\chi$
\begin{eqnarray}
&& C_0=1-2i\omega,\nonumber\\
&& C_1=-4+i(4\omega+3\chi)+\frac{i\omega^2}{\chi},\nonumber\\
&& C_2=3-i(2\omega+\chi)-\frac{i\omega^2}{\chi},\nonumber\\
&& C_3=\bigg(\frac{\omega+\chi}{\chi}\bigg)\bigg[(\omega+\chi)^2+\frac{i(\omega+3\chi)}{2}\bigg]-l(l+1)-1,\nonumber\\
&& C_4=-\bigg[\frac{(\omega+\chi)^2}{2\chi}+i\bigg]^2.
\end{eqnarray}
The boundary conditions are satisfied when the continued fraction
condition on the recursion coefficients holds. The series in
(\ref{wbess}) converge for the given $l$. The frequency $\omega$ is
a root of the continued fraction equation
\begin{eqnarray}
\beta_n-\frac{\alpha_{n-1}\gamma_n}{\beta_{n-1}-\frac{\alpha_{n-2}\gamma_{n-1}}{\beta_{n-2}
-\alpha_{n-3}\gamma_{n-2}/...}}=\frac{\alpha_{n}\gamma_{n+1}}{\beta_{n+1}-\frac{\alpha_{n+1}\gamma_{n+2}}{\beta_{n+2}
-\alpha_{n+2}\gamma_{n+3}/...}}.
\end{eqnarray}
Numerical solutions of this algebraic equation give us the
quasinormal spectrum of phantom scalar perturbations in the
Schwarzschild black string spacetime.
\TABLE{
\begin{tabular}[b]{cccccc}
 \hline \hline
$k$ & $\mu=0$ &$\mu=0.01$
 & $\mu=0.02$&  $\mu=0.03$ \\ \hline
&&&&\\
0& 0.22091-0.20979i&0.22094-0.20998i
 & 0.22101-0.21056i&
 0.22114-0.21151i
 \\
0.01&0.22088-0.20960i&0.22091-0.20979i&0.22099-0.21037i&
0.22111-0.21132i
 \\
0.02&0.22081-0.20902i&0.22083-0.20921i&0.22091-0.20979i&
0.22104-0.21075i
 \\
0.03&0.22068-0.20805i&0.22071-0.20825i&0.22078-0.20883i&
0.22091-0.20979i
\\
0.04&0.22050-0.20668i&0.22053-0.20687i&0.22060-0.20746i&
0.22073-0.20844i
\\
\hline \hline
\end{tabular}
\caption{The fundamental ($n=0$) quasinormal frequencies  of phantom
scalar field (for fixed $l=0$) in the Schwarzschild black string
spacetime.\label{1:tab}} }

\TABLE{
\begin{tabular}[b]{ccccc}
 \hline \hline
$k$&$\mu=0$ & $\mu=0.01$
 &$\mu=0.02$& $\mu=0.03$ \\ \hline
&&&& \\
0&0.58587-0.19532i&0.58581-0.19537i
 &0.58563-0.19553i&0.58534-0.19579i
 \\
0.01&0.58593-0.19527i&0.58587-0.19532i&0.58569-0.19548i&
~0.58540-0.19574i
 \\
0.02&0.58611-0.19511i&0.58605-0.19516i&0.58587-0.19532i&
~0.58558-0.19558i
 \\
0.03&0.58641-0.19485i&0.58635-0.19490i&0.58617-0.1950i&
~0.58587-0.19532i
\\
\hline \hline
\end{tabular}
\caption{The fundamental ($n=0$) quasinormal frequencies of phantom
scalar field (for fixed $l=1$) in the Schwarzschild black string
spacetime. \label{2:tab}}}

 \TABLE{
\begin{tabular}[b]{ccccc}
 \hline \hline
 $k$ &$\mu=0$&$\mu=0.01$
 &$\mu=0.02$&$\mu=0.03$ \\ \hline
&&&& \\
0&0.96729-0.19352i&0.96724-0.19354i
 &0.96711-0.19360i&
 0.96690-0.19370i
 \\
0.01&0.96733-0.19350i&0.96729-0.19352i&0.96716-0.19358i&
0.96694-0.19368i
 \\
0.02&0.96746-0.19344i&0.96742-0.19346i&0.96729-0.19352i&
0.96707-0.19362i
 \\
0.03&0.96768-0.19333i&0.96763-0.19335i&0.96750-0.19341i&
0.96729-0.19352i
\\
\hline \hline
\end{tabular}
\caption{The fundamental ($n=0$) quasinormal frequencies of phantom
scalar field (for fixed $l=2$) in the Schwarzschild black string
spacetime.\label{3:tab}} }

In tables \ref{1:tab}-\ref{3:tab}, we list the fundamental
quasinormal frequencies of phantom scalar perturbation field for
fixed $l=0\sim 2$ in the Schwarzschild black string spacetime. From
the tables \ref{1:tab}-\ref{3:tab} and figures (1)(2), one can find
that for $l=0$ with the increase of $k$ the real part first
decreases and then increases, but with the increase of $\mu$ it is
in inverse. Moreover, for the $l=1$ and $l=2$ the real part
increases with $k$ and decreases with $\mu$. Therefore the effect of
$k$ on the quasinormal frequencies is different from that of the
mass $\mu$ of phantom scalar perturbation. The main reason is that
the sign of $k^2$ in the $V(r)$ is opposite to that of $\mu^2$,
which yields that their effects on the potential are different. The
absolute value of imaginary parts for all $l$ decrease with $k$ and
increase with $\mu$. This can be explained by that the smaller $k$
and larger $\mu$ leads to the lower peak of the potential and thus
it is easier for the wave to be absorbed into the black hole.
\FIGURE[!t]{ \centerline{
\includegraphics[width=0.32\textwidth,angle=0]{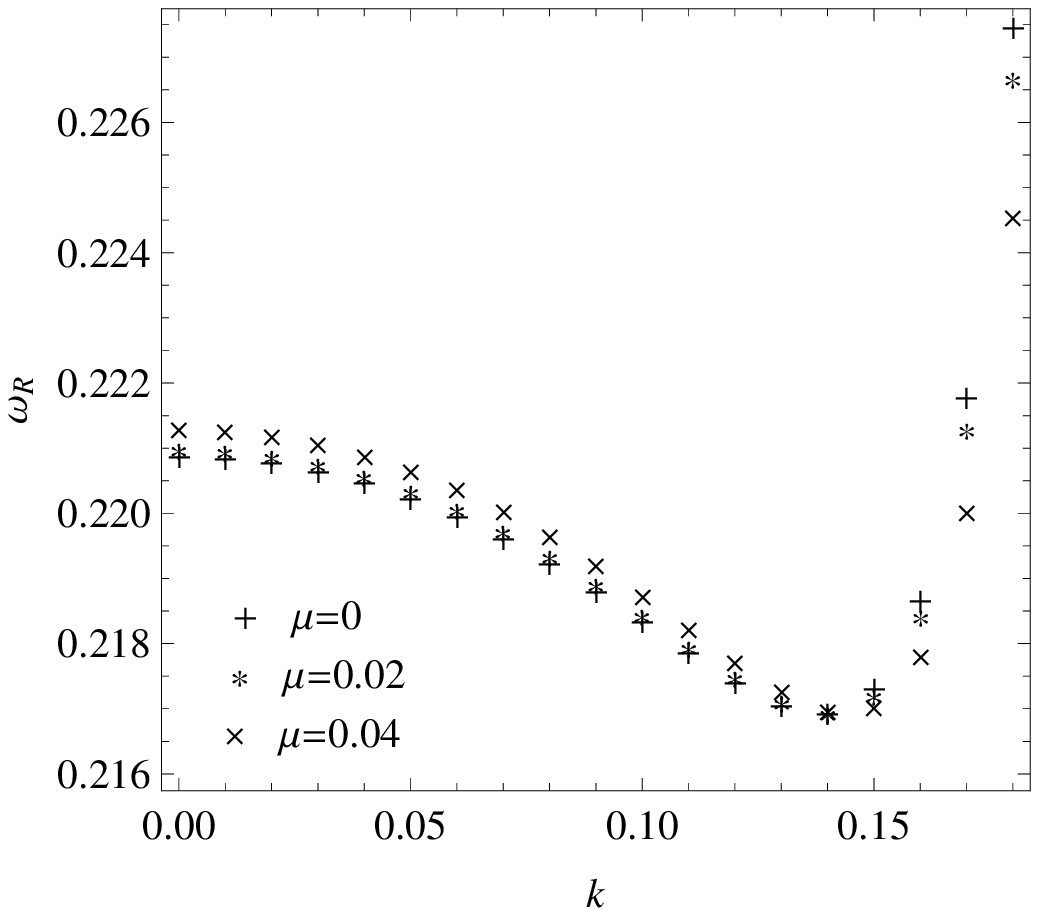}~~
\includegraphics[width=0.32\textwidth,angle=0]{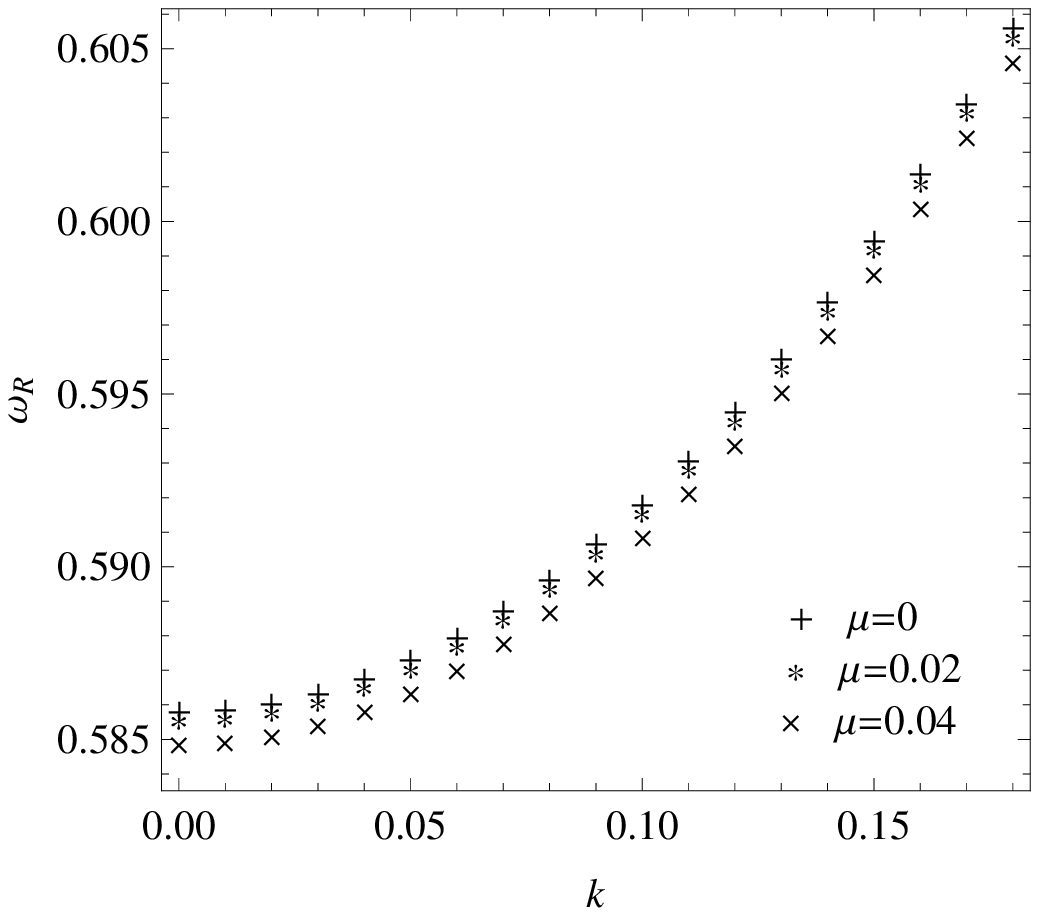}~~
\includegraphics[width=0.32\textwidth,angle=0]{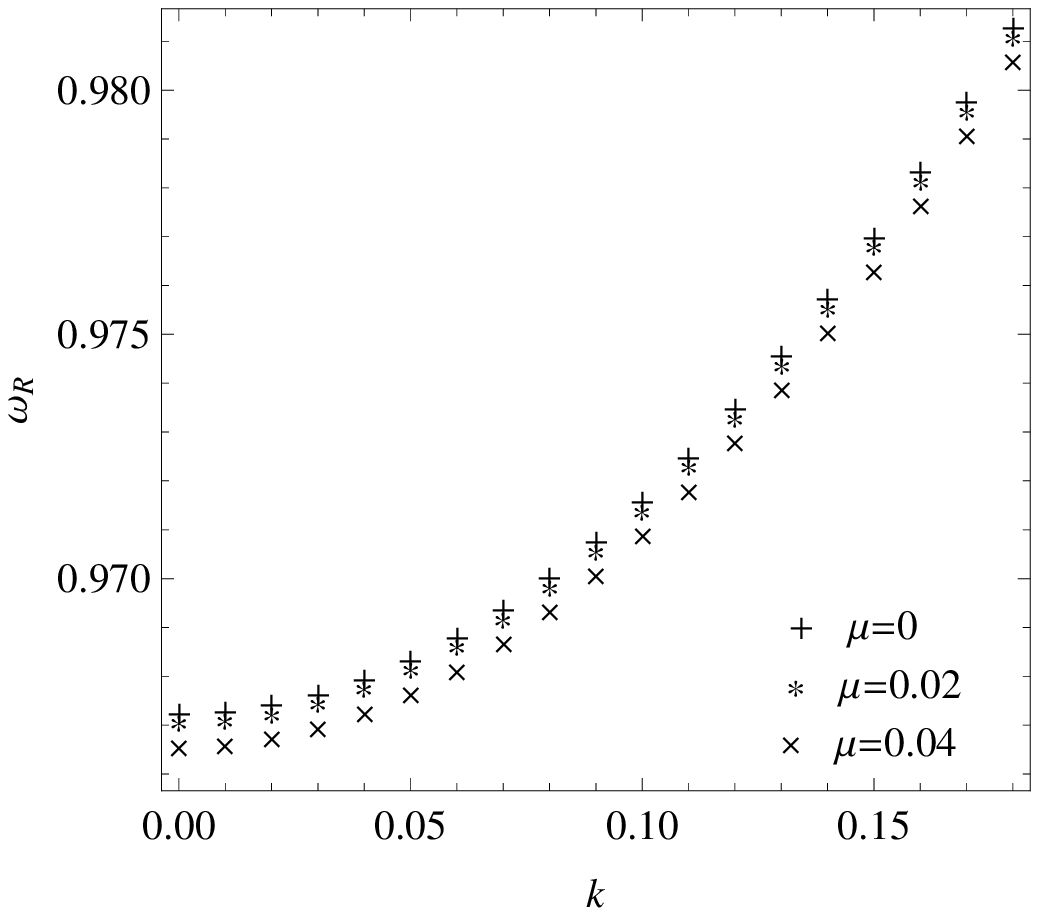} \caption{Variety of the real part quasinormal frequencies ($n=0$) for
the phantom scalar perturbations with $k$ for different $\mu$. The
left, middle and right are for $l=0$, $l=1$ and $l=2$.} }}
\FIGURE[!t]{ \centerline{
\includegraphics[width=0.32\textwidth,angle=0]{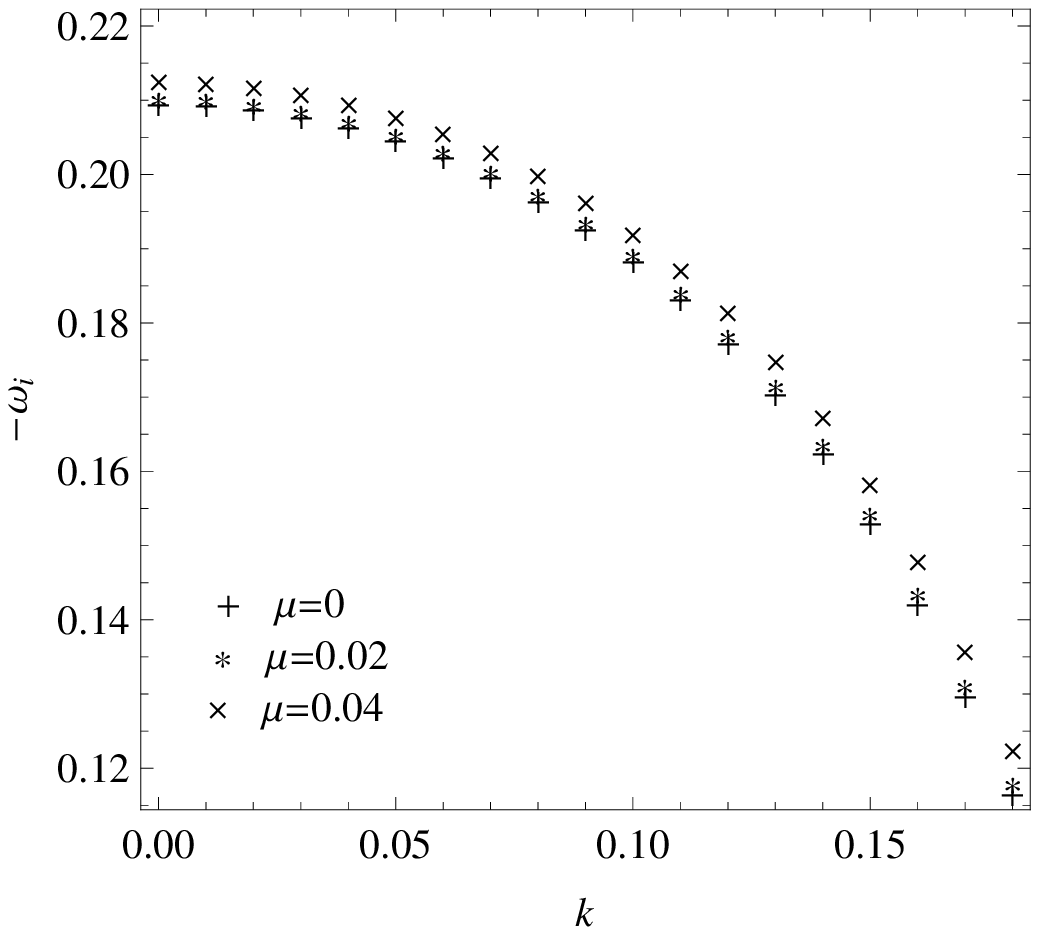}~~
\includegraphics[width=0.32\textwidth,angle=0]{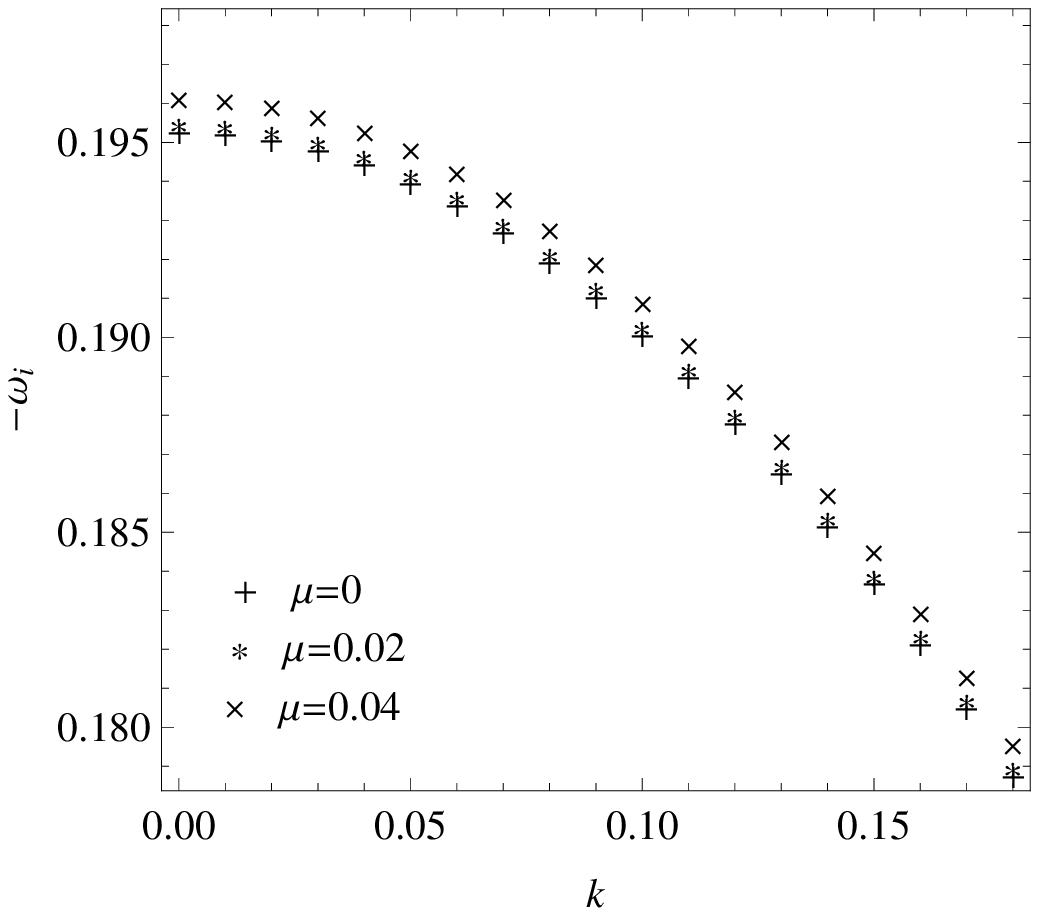}~~
\includegraphics[width=0.32\textwidth,angle=0]{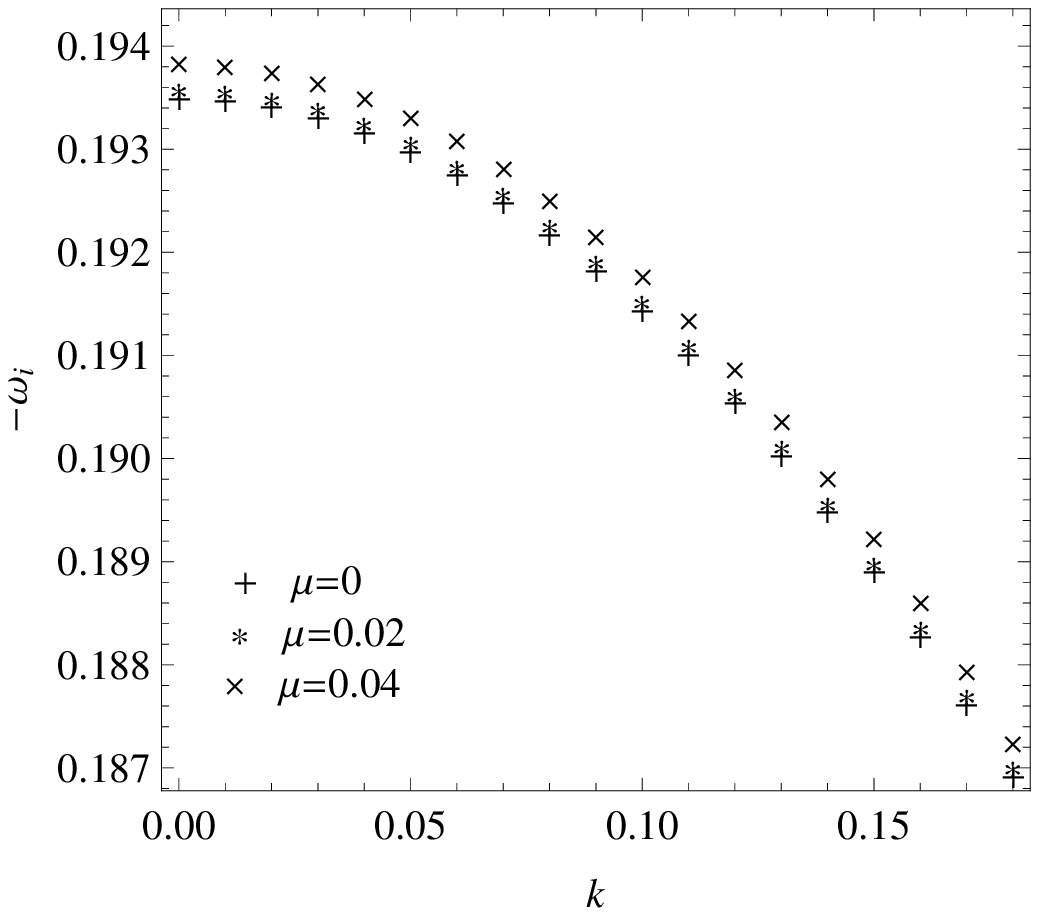} \caption{Variety of the imaginary part quasinormal frequencies ($n=0$) for
the phantom scalar perturbations with $k$ for different $\mu$. The
left, middle and right are for $l=0$, $l=1$ and $l=2$.}}}

We are now in a position to study the late-time behavior of phantom
scalar perturbations in the Schwarzschild black string spacetime by
using of the time domain method \cite{T1}. Using the light-cone
variables $u=t-r_*$ and $v=t+r_*$, the wave equation
\begin{eqnarray}
-\frac{\partial^2\psi}{\partial t^2}+\frac{\partial^2\psi}{\partial
r_*^2}=V(r)\psi,
\end{eqnarray}
can be rewritten as
 \begin{eqnarray}
4\frac{\partial^2\psi}{\partial u\partial
v}+V(r)\psi=0.\label{wbes1}
\end{eqnarray}
The two-dimensional wave equation (\ref{wbes1}) can be integrated
numerically by using the finite difference method suggested in
\cite{T1}. In terms of Taylor's theorem, it is discretized as
\begin{eqnarray}
\psi_N=\psi_E+\psi_W-\psi_S-\delta u\delta v
V(\frac{v_N+v_W-u_N-u_E}{4})\frac{\psi_W+\psi_E}{8}+O(\epsilon^4)=0,\label{wbes2}
\end{eqnarray}
where we have used the following definitions for the points: $N$:
$(u+\delta u, v+\delta v)$, $W$: $(u + \delta u, v)$, $E$: $(u, v +
\delta v)$ and $S$: $(u, v)$. The parameter $\epsilon$ is an overall
grid scalar factor, so that $\delta u\sim\delta v\sim\epsilon$.
Considering that the behavior of the wave function is not sensitive
to the choice of initial data, we set $\psi(u, v=v_0)=0$ and use a
Gaussian pulse as an initial perturbation, centered on $v_c$ and
with width $\sigma$ on $u=u_0$ as
\begin{eqnarray}
\psi(u=u_0,v)=e^{-\frac{(v-v_c)^2}{2\sigma^2}}.\label{gauss}
\end{eqnarray}
\FIGURE[!t]{
\centerline{\includegraphics[width=0.32\textwidth,angle=0]{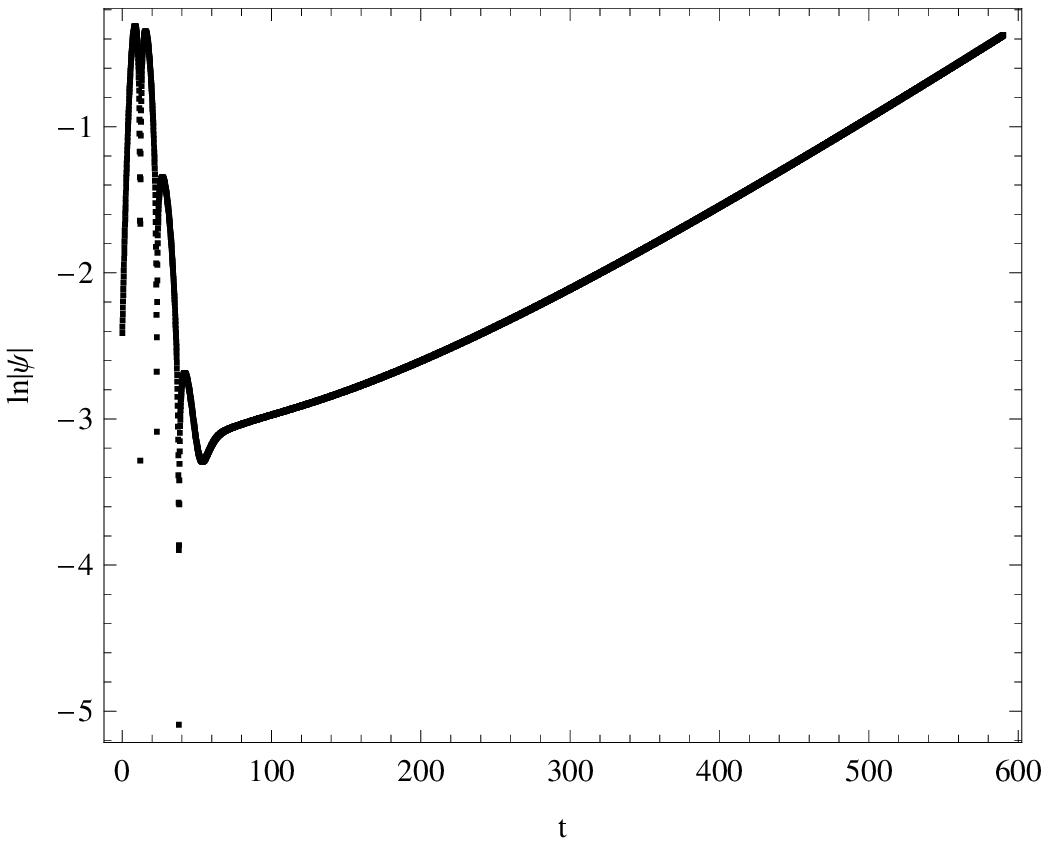}~~
\includegraphics[width=0.32\textwidth,angle=0]{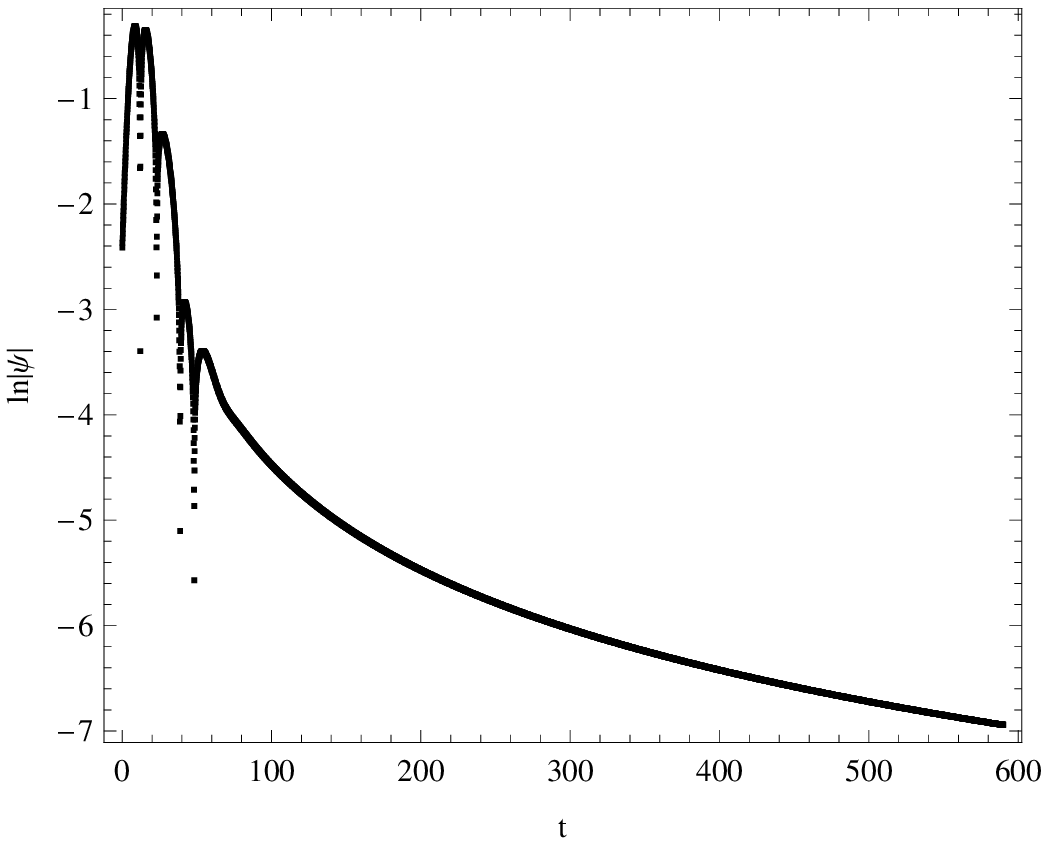}~~
\includegraphics[width=0.32\textwidth,angle=0]{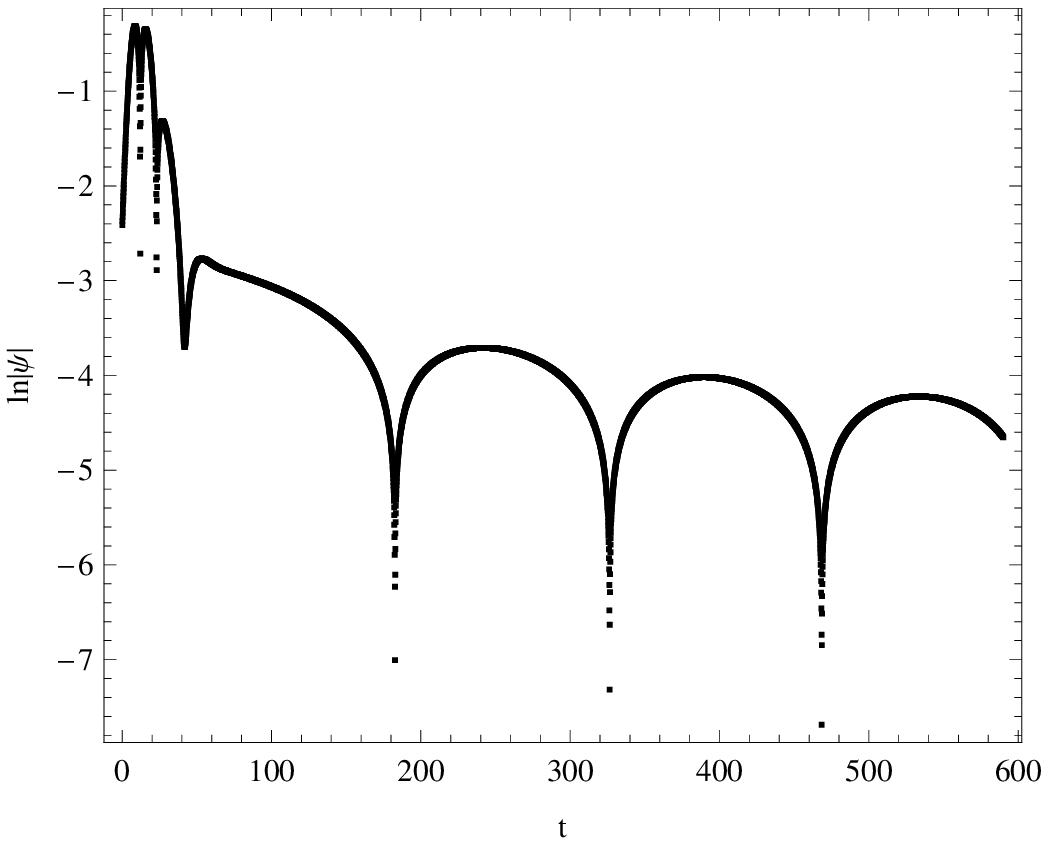} \caption{ The late-time
behaviors of the phantom scalar perturbations for fixed $l=0$ and
$\mu=0.02$, the left, middle and right are for $k=0.01$, $0.02$ and
$0.03$, respectively. The constants in the Gauss pulse (\ref{gauss})
$v_c=10$ and $\sigma=3$.} } }

\FIGURE[!t]{
\centerline{\includegraphics[width=0.32\textwidth,angle=0]{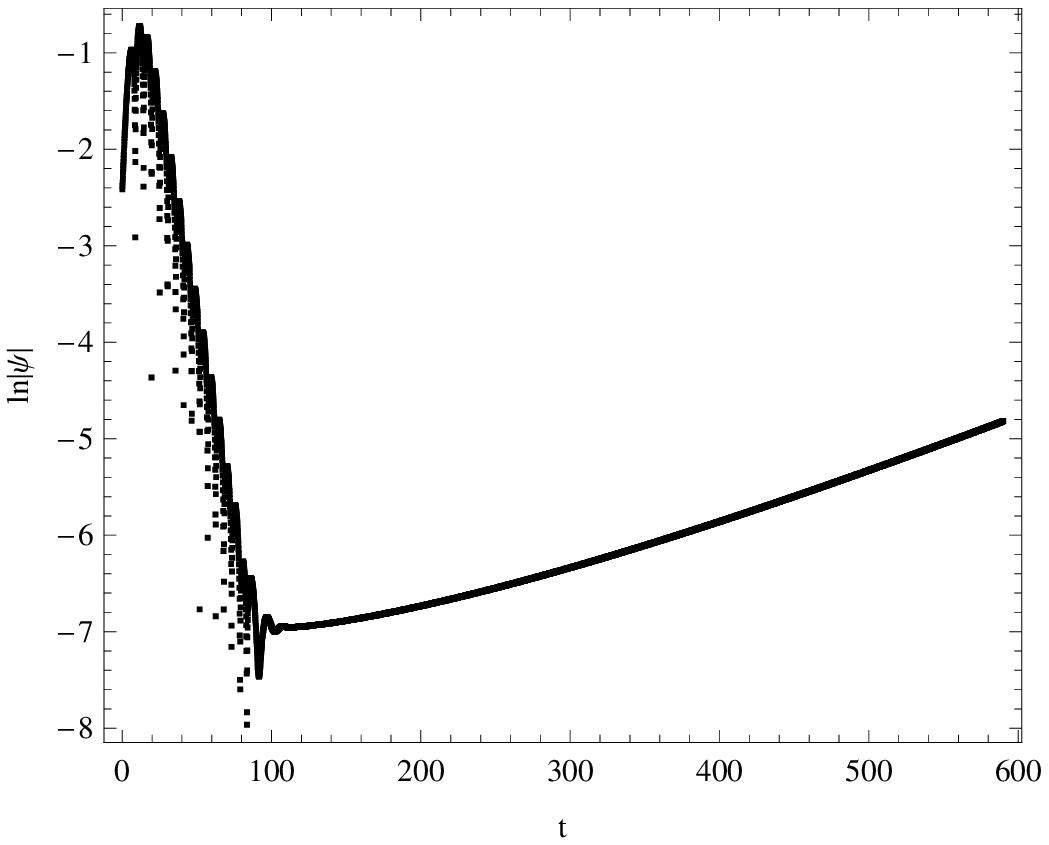}~~
\includegraphics[width=0.32\textwidth,angle=0]{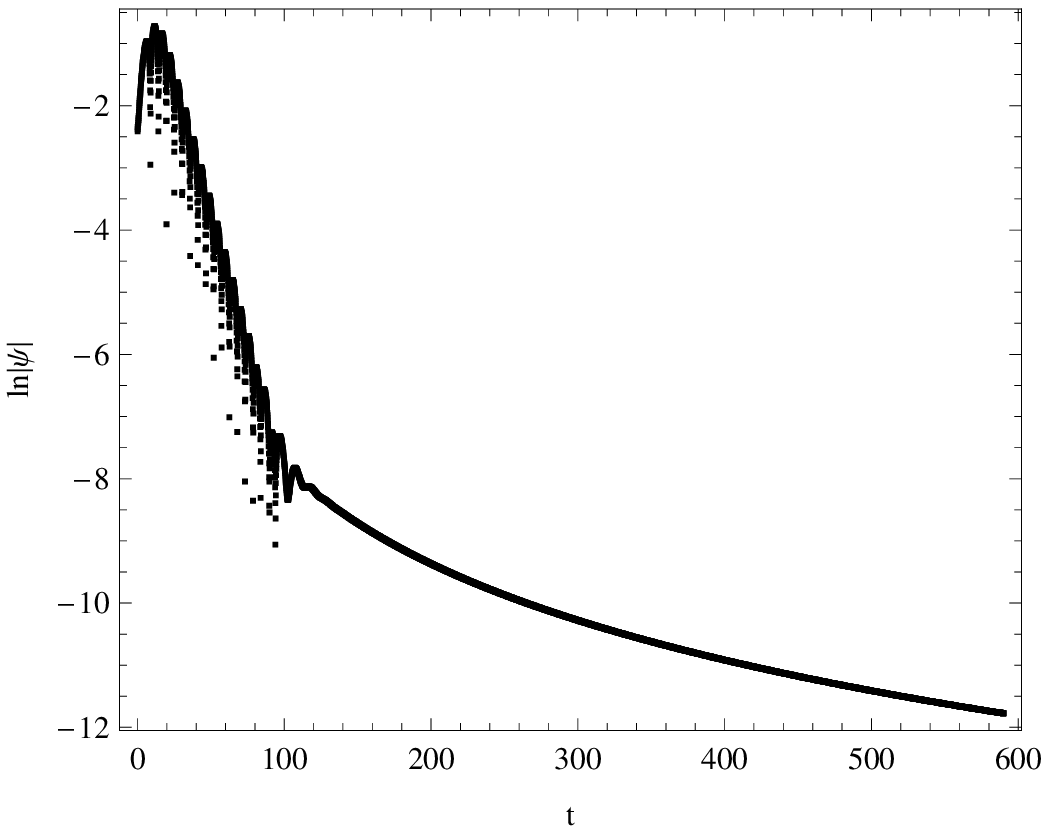}~~
\includegraphics[width=0.32\textwidth,angle=0]{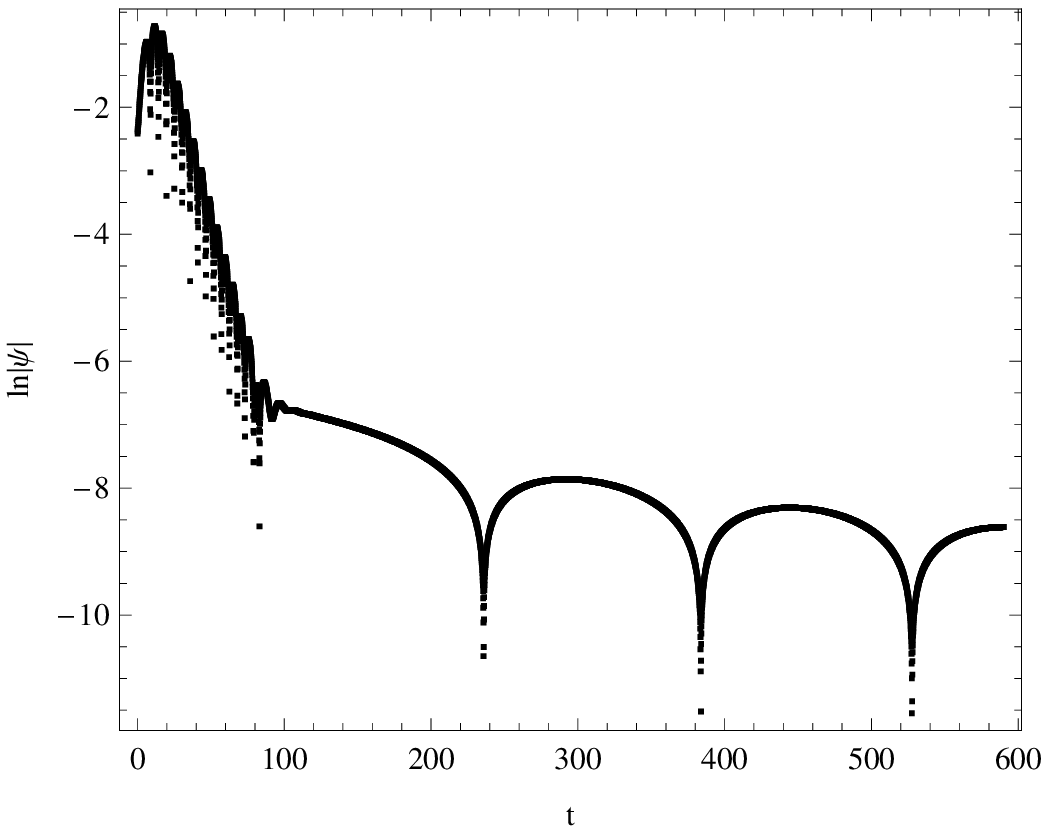}
\caption{ The late-time behaviors of the phantom scalar
perturbations for fixed $l=1$ and $\mu=0.02$, the left, middle and
right are for $k=0.01$, $0.02$ and $0.03$, respectively. The
constants in the Gauss pulse (\ref{gauss}) $v_c=10$ and
$\sigma=3$.}} }

Here we examine numerically the late-time behaviors of phantom
scalar perturbations in the Schwarzschild black string spacetime. In
figures (4) and (5), we plot the evolutions of the phantom scalar
perturbations for fixed $l=0$ and $l=1$ respectively. For all $l$,
the late-time behaviors of phantom scalar perturbation depend
heavily on the sign of the quantity $k^2-\mu^2$. When it is negative
(i.e. $k^2<\mu^2$), we find the phantom scalar field after
undergoing the quasinormal modes grow with exponential rate. As in
the Schwarzschild black hole spacetime, the asymptotic behaviors of
the wave function have the form
\begin{eqnarray}
\psi\sim e^{\alpha \sqrt{\mu^2-k^2} t-4l-\beta},
\end{eqnarray}
where $\alpha$ and $\beta$ are two numerical constant. This behavior
can be attributed to that the effective potential is negative at the
spatial infinity and the wave outside the black hole gains energy
from the spacetime \cite{bwr}. Moreover, the exponential growth of
the wave function means also that in this case the Schwarzschild
black string spacetime is unstable against the external phantom
scalar perturbation. When the wavenumber $k$ is equal to the mass
$\mu$ of the phantom scalar perturbation, we find that it decays
without any oscillation, which is similar to that of usual massless
scalar field in the Schwarzschild black hole spacetimes
\cite{T1,RH1, RH2,RH3}. It is not surprising because that in this
case the effective potential (\ref{efp}) has the same form as that
of the usual massless scalar field. When the quantity $k^2-\mu^2$ is
positive, the figures (4) and (5) tell us the behaviors of the
phantom scalar perturbation have the form
$t^{-\gamma}\sin{(\sqrt{k^2-\mu^2}~t)}$. This implies that the
phantom scalar field decays with the oscillatory inverse power-law
behavior which is similar to that of the usual massive scalar
perturbations in a black hole spacetime \cite{ml1,ml2}. These
properties of phantom scalar field have not been observed elsewhere.

From the discussion above, we know that in the case $k^2\geq\mu^2$
the Schwarzschild black string spacetime is stable when it suffers
from the external phantom-like perturbations. Thus there exists a
threshold values of the wavenumber $k$ at which the stability of
phantom perturbations appears. In other words,  only the phantom
scalar perturbations whose wavelengths satisfy $\lambda>2\pi/\mu$ is
unstable in Schwarzschild black string spacetime. For the smaller
$\mu$, the range of the wavelength $\lambda$ of the phantom field is
stable becomes wider. For the fixed $\mu$, only the phantom
perturbation has the longer wavelength is unstable. This result is
consistent with that of the Gergory-Laflamme instability which is
obtained in the evolution of the scalar type of the gravitational
perturbation in this background. For the usual scalar perturbations,
the quantity $k^2-\mu^2$ in the effective potential $V(r)$
(\ref{efp}) is replaced by $k^2+\mu^2$. Thus for the usual scalar
perturbations there does not exist such threshold value for the
wavenumber $k$ and the perturbations are always stable in the
Schwarzschild black string spacetime.

\section{Summary}

In this paper we examined the dynamical evolution of the phantom
scalar perturbation in the Schwarzschild black string spacetime. We
find that the quasinormal modes and late-time behaviors contain the
imprint from the fifth dimension. The form of the late-time
evolution is determined by the difference between the wavenumber $k$
and the mass $\mu$ of the phantom scalar perturbation. For $k<\mu$,
the phantom scalar perturbation in the late-time evolution grows
with an exponential rate as in the four-dimensional Schwarzschild
black hole spacetime. While, for $k=\mu$, the late-time behavior has
the same form as that of the massless scalar field perturbation in
the background of a black hole. Furthermore, for $k>\mu$, the
late-time evolution of phantom scalar perturbation is dominated by a
decaying tail with an oscillation which is consistent with that of
the usual massive scalar field. These information can help us know
more about the wave dynamics of phantom scalar perturbation and the
properties of black string. It would be of interest to generalize
our study to other spacetimes. Work in this direction will be
reported in the future.

\acknowledgments
This work was partially supported by the National
Natural Science Foundation of China under Grant No.10875041; the
Scientific Research Fund of Hunan Provincial Education Department
Grant No.07B043 and the construct program of key disciplines in
Hunan Province. J. L. Jing's work was partially supported by the
National Natural Science Foundation of China under Grant No.10675045
and No.10875040; and the Hunan Provincial Natural Science Foundation
of China under Grant No.08JJ3010.


\begin{thebibliography}{}
\bibitem{1a} S. Weinberg, \textit{The cosmological constant problem}, Rev. Mod. Phys. {\bf61} (1989) 1;

V. Sahni and A. Starobinsky, \textit{The Case for a Positive
Cosmological Lambda-term} Int. J. Mod. Phy. D. {\bf9} (2000) 373;

P. J. E. Peebles and B. Ratra, \textit{The cosmological constant and
dark energy}, Rev. Mod. Phys. {\bf75}, 559 (2003);

T. Padmanabhan, \textit{Cosmological Constant-the Weight of the
Vacuum}, Phys. Rept. {\bf 380} (2003) 235.

\bibitem{2a} B. Ratra and P. J. E. Peebles, \textit{Cosmological consequences of a rolling homogeneous scalar
field,} Phys. Rev. D {\bf 37} (1988) 3406;

P. J. E. Peebles  and B. Ratra, \textit{Cosmology with a
Time-Variable Cosmological ``Constant"}, Astrophys. J. Lett. {\bf
325} (1988) L17;

C. Wetterich, \textit{Cosmology and the fate of dilatation
symmetry},  Nucl. Phys. B {\bf 302} (1988) 668;

R. R. Caldwell, R. Dave and P. J. Steinhardt, \textit{Cosmological
imprint of an energy component with general equation-of-state},
Phys. Rev. Lett. {\bf 80} (1998) 1582;

I. Zlatev, L. Wang and P. J. Steinhardt, \textit{Quintessence,
Cosmic Coincidence, and the Cosmological Constant},  Phys. Rev.
Lett. {\bf 82} (1999) 896;

M. Doran and J. Jaeckel, \textit{ Loop Corrections to Scalar
Quintessence Potentials}, Phys. Rev. D. {\bf 66} (2002) 043519 .

\bibitem{3a} C. A.  Picon, T. Damour and V. Mukhanov, \textit{k-Inflation}, Phys. Lett. B
{\bf458} (1999) 209;

T. Chiba, T. Okabe and M. Yamaguchi, \textit{Kinetically driven
quintessence}, Phys. Rev. D {\bf62} (2000) 023511.

\bibitem{4a} R. R. Caldwell, \textit{A phantom menace? Cosmological consequences of a dark energy component with super-negative equation of state},
Phys. Lett. B {\bf 545} (2002) 23;

B. McInnes, \textit{The dS/CFT Correspondence and the Big Smash},
JHEP {\bf 08} (2002) 029;

S. Nojiri and S. D. Odintsov, \textit{Bulk and brane gauge
propagator on 5d AdS black hole}, Phys. Lett. B {\bf 562} (2003)
147;

L. P. Chimento and R. Lazkoz, \textit{Constructing Phantom
Cosmologies from Standard Scalar Field Universes}, Phys. Rev. Lett.
{\bf 91} (2003) 211301;

B. Boisseau, G. Esposito-Farese, D. Polarski, Alexei A. Starobinsky,
\textit{Reconstruction of a Scalar-Tensor Theory of Gravity in an
Accelerating Universe}, Phys. Rev. Lett. {\bf 85} (2000) 2236;

R. Gannouji, D. Polarski, A. Ranquet, A. A. Starobinsky, \textit{
Scalar-Tensor Models of Normal and Phantom Dark Energy}, JCAP {\bf
0609} (2006) 016, [astro-ph/0606287].

\bibitem{41a} P. Candelas, \textit{Vacuum polarization in Schwarzschild spacetime}, Phys. Rev. D {\bf 21} (1980) 2185;

D. W. Sciama, P. Candelas and D. Deutsch, \textit{Quantum Field
Theory, Horizons and Thermodynamics}, Adv. Phys. {\bf 30} (1981)
327.

\bibitem{42a} B. McInnes, \textit{The Strong Energy Condition and the S-Brane Singularity Problem}, JHEP {\bf0306}
 (2003) 043,  hep-th/0305107;

M. N. R. Wohlfarth, \textit{ Inflationary Cosmologies from
Compactification?}, Phys. Rev. D {\bf 69} (2004) 066002,
hep-th/0307179;

N. Ohta, \textit{Accelerating Cosmologies from S-Branes}, Phys. Rev.
Lett. {\bf 91} (2003) 061303, hep-th/0303238; \textit{A Study of
Accelerating Cosmologies from Superstring/M theories}, Prog. Theor.
Phys. {\bf 110} (2003) 269, hep-th/0304172;

S. Roy, \textit{ Accelerating cosmologies from M/String theory
compactifications}, Phys. Lett. B {\bf 567} (2003) 322,
hep-th/0304084;

M. Gutperle, R. Kallosh and A. Linde, \textit{ M/String Theory,
S-branes and Accelerating Universe}, JCAP {\bf 0307} (2003) 001,
hep-th/0304225;

C. M. Chen, P. Ho, I. P. Neupane and J. E. Wang, \textit{ A Note on
Acceleration from Product Space Compactification}, JHEP {\bf 0307}
(2003) 017, hep-th/0304177.

C. M. Chen, P. Ho, I. P. Neupane, N. Ohta and J. E. Wang,
\textit{Hyperbolic Space Cosmologies}, JHEP {\bf 0310} (2003) 058,
hep-th/0306291.

Ishwaree P. Neupane, \textit{Accelerating Cosmologies from
Exponential Potentials}, Class. Quant. Grav. {\bf 21} (2004) 4383,
hep-th/0311071.

\bibitem{6a1} R. R. Caldwell, M. Kamionkowski, and N. N. Weinberg, \textit{Phantom Energy: Dark Energy with $w<-1$ Causes a Cosmic Doomsday}, Phys. Rev. Lett.
 {\bf91} (2003) 071301;

S. Nesseris and L. Perivolaropoulos, \textit{Fate of bound systems
in phantom and quintessence cosmologies}, Phys. Rev. D {\bf 70}
(2004) 123529;

S. Nojiri and S. D. Odintsov, \textit{Effective equation of state
and energy conditions in phantom/tachyon inflationary cosmology
perturbed by quantum effects}, Phys. Lett. B {\bf 571} (2003) 1.

\bibitem{6a11} P. Singh, M. Sami and N. Dadhich,  \textit{Cosmological dynamics of a phantom field}, Phys. Rev. D {\bf 68} (2003) 023522.

\bibitem{6a2} J. G. Hao and X. Z. Li, \textit{Phantom cosmic dynamics: Tracking attractor and cosmic doomsday}, Phys. Rev. D {\bf 70} (2004) 043529.

\bibitem{6a3}  L. A. Urena-Lopez, \textit{Scalar phantom energy as a cosmological dynamical system}, JCAP {\bf 0509} (2005) 013.

\bibitem{6a4} S. B. Chen, B. Wang ang J. L. Jing, \textit{Dynamics of an interacting dark energy model in Einstein and loop quantum cosmology},
Phys. Rev. D {\bf 78} (2008) 123503.

\bibitem{ta}P. F. Gonzalez-Diaz and C. L. Siguenza, \textit{Phantom thermodynamics},  Nucl. Phys B {\bf
697} (2004) 363;

I. Brevik, S. Nojiri, S. D. Odintsov, and L. Vanzo, \textit{Entropy
and universality of the Cardy-Verlinde formula in a dark energy
universe}, Phys. Rev. D {\bf 70} (2004) 043520;

S. H. Pereira, J. A. S. Lima , \textit{On Phantom Thermodynamics},
Phys. Lett. B {\bf 669} (2008) 266;

\bibitem{5a} A. Melchiorri, L. Mersini-Houghton, C. J. Odman,
and M. Trodden, \textit{The state of the dark energy equation of
state}, Phys. Rev. D {\bf68} (2003) 043509.

\bibitem{Eph} E. Babichev, V. Dokuchaev and Y. Eroshenko, \textit{Black Hole Mass Decreasing due to Phantom Energy Accretion},
 Phys. Rev. Lett. {\bf93} (2004) 021102.

\bibitem{Eph1} E. Babichev, S. Chernov, V. Dokuchaev and Y. Eroshenko, \textit{Phantom threat to cosmic censorship}, arXiv:0806.0916

\bibitem{Eph2} M. Jamil, M. A. Rashid and A. Qadir, \textit{Charged Black Holes in Phantom Cosmology}, Eur.
Phys. J. C {\bf58}  (2008) 325.

\bibitem{sb1} S. B. Chen, J. L. Jing and Q. Y. Pan, \textit{Wave dynamics of phantom scalar perturbation in the background of Schwarzschild black hole
}, Phys. Lett. B {\bf670} (2009) 276, arXiv:0809.1152.

\bibitem{qn1} H. P. Nollert, \textit{Quasinormal modes: the characteristic ``sound" of black holes and neutron stars},
Class. Quantum Grav. {\bf 16} (1999) R159.

\bibitem{qn2} K. D. Kokkotas, B. G. Schmidt, \textit{Quasi-normal modes of stars and black holes}, Living Rev. Rel. {\bf 2} (1999) 2.

\bibitem{qn3} B. Wang, \textit{Perturbations around black holes}, Braz. J. Phys. {\bf35} (2005) 1029.

\bibitem{ml1}H. Koyama and A. Tomimatsu, \textit{Asymptotic power-law tails of massive scalar fields in a Reissner-Nordstr?m background},
Phys. Rev. D {\bf 63} (2001) 064032;

H. Koyama and A. Tomimatsu, \textit{Asymptotic tails of massive
scalar fields in a Schwarzschild background}, Phys. Rev. D {\bf 64}
(2001) 044014;

R. Moderski and M. Rogatko, \textit{Late-time evolution of a
self-interacting scalar field in the spacetime of a dilaton black
hole }, Phys. Rev. D {\bf64} (2001) 044024;

R. Moderski and M. Rogatko, \textit{Evolution of a self-interacting
scalar field in the spacetime of a higher dimensional black hole},
Phys. Rev. D {\bf 72} (2005) 044027;

S. Hod and T. Piran, \textit{Late-time tails in gravitational
collapse of a self-interacting (massive) scalar-field and decay of a
self-interacting scalar hair}, Phys. Rev. D {\bf 58} (1998) 044018;

S. B. Chen and J. L. Jing, \textit{Late-time tail of a coupled
scalar field in the background of a black hole with a global
monopole }, Mod. Phys. Lett. A {\bf 23} (2008) 35 ;

S. B. Chen, B. Wang and R. K. Su, \textit{ Quasinormal modes and
late-time tails in the background of Schwarzschild black hole
pierced by a cosmic string: scalar, electromagnetic and
gravitational perturbations }, Int. J. Mod. Phys. A {\bf 16} (2008)
2502.

\bibitem{ml2} S. Hod, \textit{Late-time evolution of realistic rotating collapse and the no-hair theorem}, Phys. Rev. D {\bf 58} (1998) 104022;

 L. Barack and A. Ori, \textit{Late-Time Decay of Scalar Perturbations Outside Rotating Black Holes}, Phys. Rev. Lett. {\bf 82} (1999) 4388;

 W. krivan, \textit{Late-time dynamics of scalar fields on rotating black hole backgrounds}, Phys. Rev. D {\bf 60} (1999) 101501(R);

\bibitem{sb2} S. B. Chen and J. L. Jing, \textit{Phantom scalar emission in the Kerr black hole spacetime}, arXiv: 0809.3164.

\bibitem{wmh} V. P. Frolov and I. D. Novikov, \textit{Black Hole Physics: Basic
Concepts and New Developments}, (Kluwer, Dordrecht, 1998);

S. V. Sushkov, \textit{Wormholes supported by a phantom energy},
Phys. Rev. D{\bf 71}, (2005) 043520 ;

O. B. Zaslavskii, \textit{Exactly solvable model of wormhole
supported by phantom energy},  Phys. Rev. D{\bf 72}, (2005) 061303;

F.Rahaman, M. Kalam, M. Sarker, K. Gayen, \textit{A theoretical
construction of wormhole supported by Phantom Energy},  Phys. Lett.
B{\bf 633}, (2006) 161-163;

M. Cataldo, P. Labrana, S. Campo, J. Crisostomo and P. Salgado,
\textit{Evolving Lorentzian wormholes supported by phantom matter
with constant state parameters}, Phys. Rev. D{\bf 78}, (2008)
104006.

\bibitem{GF1} R. Gregory and R. Laflamme, \textit{Black strings and p-branes are unstable
}, Phys. Rev. Lett. {\bf 70} 2837 (1993), arXiv:hep-th/9301052.

\bibitem{GF2} R. Gregory and R. Laflamme, \textit{The Instability of Charged Black Strings and p-Branes}, Nucl. Phys. B {\bf 428}, 399 (1994), arXiv:hep-th/9404071.
\bibitem{GF3} T. Harmark, V. Niarchos and N. A. Obers, \textit{Instabilities of Black Strings and Branes}, Class. Quant. Grav. {\bf 24}, R1 (2007),
arXiv:hep-th/0701022.
\bibitem{GF4} R. A. Konoplya, K. Murata, Jiro Soda and A. Zhidenko, \textit{Looking at the Gregory-Laflamme instability through quasinormal modes},
Phys. Rev. D {\bf 78} (2008) 084012.

\bibitem{GF5} J. L. Hovdebo and R. C. Myers, \textit{Black rings, boosted strings, and Gregory-Laflamme instability}, Phys. Rev. D {\bf 73} (2006)
084013, arXiv:hep-th/0601079.


\bibitem{Leaver}E. W. Leaver, \textit{An analytic representation for the quasi-normal modes of Kerr black holes}, Proc. R. Soc. Lond.
A {\bf 402} (1985) 285;

E. W. Leaver, \textit{Spectral decomposition of the perturbation
response of the Schwarzschild geometry}, Phys. Rev. D {\bf 34}
(1986) 384.

\bibitem {T1} C. Gundlach, R. H. Price and J. Pullin, \textit{Late-time behavior of stellar collapse and explosions. I. Linearized perturbations},
Phys. Rev. D {\bf 49} (1994) 883.

\bibitem{phso1}V. Dzhunushaliev, V. Folomeev, R. Myrzakulov and K. Douglas Singleton, \textit{Non-singular solutions to Einstein-Klein-Gordon equations with a phantom scalar field},
JHEP {\bf 0807} (2008) 094, arxiv: 0805.3211.

\bibitem{bwr} B. Wang, E. Abdalla and R. B Mann, \textit{Scalar wave propagation in topological black hole backgrounds
}, Phys. Rev. D {\bf 65} (2002) 084006.

\bibitem{RH1} R. H. Price, \textit{Nonspherical Perturbations of Relativistic Gravitational Collapse. I. Scalar and Gravitational Perturbations},
Phys. Rev. D {\bf 5} (1972) 2419.
\bibitem{RH2} S. Hod and T. Piran, \textit{Late-time evolution of charged gravitational collapse and decay of charged scalar hair},
Phys. Rev. D {\bf 58} (1998) 024017;

L. Barack, \textit{Late time decay of scalar, electromagnetic, and
gravitational perturbations outside rotating black holes}, Phys.
Rev. D {\bf 61} (2000) 024026;

L.M. Burko and G. Khanna, \textit{Radiative falloff in the
background of rotating black holes}, Phys. Rev. D {\bf 67} (2003)
081502;

L.M. Burko and G. Khanna, \textit{Universality of massive scalar
field late-time tails in black-hole spacetimes}, Phys. Rev. D {\bf
70} (2004) 044018.

\bibitem{RH3} E. S. C. Ching, P. T. Leung, W. M. Suen, and K. Young, \textit{Wave propagation in gravitational systems: Late time behavior},
Phys. Rev. D {\bf 52} (1995) 2118;

V. Cardoso, S. Yoshida, O. J. C. Dias, and J.P.S. Lemos,
\textit{Late-time tails of wave propagation in higher dimensional
spacetimes}, Phys. Rev. D {\bf 68} (2003) 061503(R).
\end{thebibliography}
\end{document}